\documentclass[10pt,conference]{IEEEtran}\IEEEoverridecommandlockouts
\usepackage{epsfig,graphics,subfigure,psfrag,amsmath,cases}
\usepackage{latexsym,amssymb,amsmath,epsfig,subfigure,algorithm}
\usepackage{algorithmic}
\usepackage{color}
\usepackage{url}
\usepackage{scrtime}

\author{\authorblockN{ Derrick Wing Kwan Ng\authorrefmark{1}, Ernest S. Lo\authorrefmark{2}, and Robert Schober\authorrefmark{1}\thanks{This work was supported in part by the AvH Professorship Program of the Alexander von Humboldt Foundation.}}
\authorrefmark{1}Institute for Digital Communications, Universit\"at Erlangen-N\"urnberg, Germany\\
\authorrefmark{2}Centre Tecnol\`{o}gic de Telecomunicacions de Catalunya - Hong Kong (CTTC-HK) \vspace*{-4mm}
}

\title{\vspace*{-2mm}Energy-Efficient Resource Allocation in Multiuser OFDM Systems with Wireless Information and Power Transfer\vspace*{-1mm}}

\date{\thistime,\,\today}

\newtheorem{Thm}{Theorem}

\newcommand{\abs}[1]{\lvert#1\rvert}

\begin{document}

\maketitle

\begin{abstract}
In this paper, we study the resource allocation algorithm design for multiuser orthogonal frequency division multiplexing
 (OFDM) downlink systems with simultaneous wireless information and power transfer.
The algorithm design is formulated as a non-convex optimization problem for maximizing
the energy efficiency of data transmission (bit/Joule delivered to
the users). In particular, the problem formulation takes into account the minimum required system data rate, heterogeneous minimum   required power transfers to the users, and the circuit
power consumption. Subsequently, by exploiting the method of time-sharing  and  the properties of
nonlinear fractional programming, the considered non-convex optimization
problem is solved using an efficient iterative resource allocation algorithm. For each iteration, the optimal power allocation and user selection solution are derived based on Lagrange dual decomposition.
 Simulation results illustrate that the proposed
iterative resource allocation algorithm achieves the maximum energy efficiency of the system and reveal how energy efficiency, system capacity, and wireless power transfer benefit from the presence of multiple users in the system.
\end{abstract}

\renewcommand{\baselinestretch}{0.93}
\large\normalsize

\section{Introduction}
\label{sect1}
Orthogonal frequency
division multiplexing (OFDM) is one of the leading candidates for
supporting high data rate wireless  broadband communication systems, as envisioned e.g. in the 3GPP Long Term
Evolution Advanced (LTE-A) and IEEE 802.11 a/g/n Wireless Fidelity (Wi-Fi) standards, due to its flexibility in resource allocation and ability in exploiting multiuser diversity. In practice, a wireless communication system is expected to support multiple  mobile users and to guarantee  quality of service. However, because of the limited radio resources and  harsh wireless channel conditions, some of the mobile users are typically switched to idle mode  since they cannot be served by the system temporarily. Unfortunately,   mobile devices are often battery driven and energy is dissipated even if they are idle which creates bottlenecks in perpetuating the  network's lifetime.

Recently,  driven by environmental concerns, green  mobile communication has received considerable
interest from both industry and academia \cite{JR:Mag_green}-\nocite{CN:static_power}\cite{JR:TWC_large_antennas}.
A promising approach to enhance  the energy efficiency  (bit-per-Joule) of wireless communication systems is to harvest energy from the environment.  Solar, wind, and geothermal are the major renewable  energy sources for generating electricity. However, these conventional natural energy sources may not be suitable for mobile devices and  not be available in  enclosed/indoor environments.
  On the other hand, wireless power transfer, in which energy is harvested from propagating electromagnetic  waves (EM) in radio frequency (RF), is becoming a new paradigm in energy harvesting since it recycles  the abundant ambient   RF energy \cite{CN:bio_WIP}--\nocite{JR:RFID,CN:WIPT_fundamental,CN:Shannon_meets_tesla,CN:MIMO_WIPT}\cite{JR:WIP}. Although the development of wireless power transfer technology is still in its infancy, there have been some preliminary applications of wireless power transfer such as wireless body area networks (WBAN) for biomedical implants \cite{CN:bio_WIP} and passive radio-frequency identification (RFID) systems \cite{JR:RFID}. Indeed, EM waves can carry both  information and power/energy simultaneously  \cite{CN:WIPT_fundamental}--\nocite{CN:Shannon_meets_tesla,CN:MIMO_WIPT}\cite{JR:WIP}. The utilization of this characteristic of EM waves imposes  many new challenges for wireless communication engineers. In \cite{CN:WIPT_fundamental} and \cite{CN:Shannon_meets_tesla}, the fundamental  trade-off between system capacity and wireless power transfer was studied for flat fading and frequency selective channels, respectively. However, \cite{CN:WIPT_fundamental} and \cite{CN:Shannon_meets_tesla} assumed a theoretical
receiver, which is able to decode information and extract power from the same received signal but is not yet available in practice. In \cite{CN:MIMO_WIPT} and  \cite{JR:WIP}, the authors proposed  different power allocation schemes for  multiple antenna two-user   narrowband  systems by separating the process of  information decoding and energy harvesting into two receivers. However,  if a multicarrier system with an arbitrary number of users is considered, the results in \cite{CN:MIMO_WIPT} and \cite{JR:WIP} which are valid for single-carrier transmission, may no longer be applicable.  Besides, the energy efficiency of wireless information and power transfer systems is still unknown since the power dissipations in RF transmission  and electronic circuitries have not been taken into account in the literature, e.g. \cite{CN:WIPT_fundamental}--\nocite{CN:Shannon_meets_tesla,CN:MIMO_WIPT}\cite{JR:WIP}.

In this paper, we address the above issues and focus on the resource allocation algorithm design for energy efficient
communication in multiuser OFDM systems with wireless information and power transfer.
In Section \ref{sect:OFDMA_AF_network_model}, we introduce the adopted multiuser OFDM channel model. In Section \ref{sect:forumlation}, we formulate the resource allocation algorithm design as a non-convex optimization
problem, which is solved by an efficient iterative resource allocation algorithm in Section \ref{sect:solution}. Section \ref{sect:result-discussion}
presents simulation results for the system performance, and in Section \ref{sect:conclusion}, we
conclude with a brief summary of our results.

\section{System Model}
\label{sect:OFDMA_AF_network_model}
In this section, we introduce the OFDM system model.
%%%%%%%%%%%%%%%%%%%%%%%%%%%%%%%
\subsection{Multiuser OFDM Channel Model}
\begin{figure} \vspace*{-2mm}
 \centering
\includegraphics[width=2.20 in]{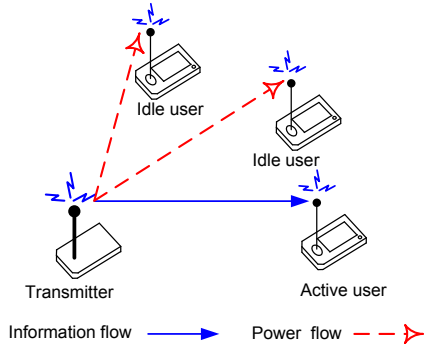}
 \caption{Illustration of a multiuser OFDM system with  $K = 3$ users for downlink wireless information and power transfer.} \vspace*{-4mm}\label{fig:system_model}
\end{figure}

%%%%%%%%%%%%%%%%%%%%%%%%%%%%%%%%
We consider a multiuser OFDM system which consists of a transmitter and $K$ mobile users. All transceivers are equipped with a
single antenna, cf.  Figure \ref{fig:system_model}. The total
 bandwidth of the system is $\cal B$ Hertz and there are $n_F$
subcarriers. Each subcarrier has a bandwidth $W={\cal B}/n_F$ Hertz.
We assume that the OFDM signaling is time slotted and the
length of each time (/scheduling) slot is comparable to the length of the channel coherence time; the channel impulse
response is assumed to be time invariant during each scheduling slot. As a result, the
downlink channel state information (CSI) can be accurately obtained by exploiting feedback from users in frequency division duplex (FDD) systems and channel reciprocity in time division duplex (TDD) systems. At  the beginning of each
scheduling slot, the transmitter computes the resource allocation policy based on the available CSI. The downlink
received symbol at user $k\in\{1,\,\ldots,\,K\}$
on subcarrier $i\in\{1,\,\ldots,\,n_F\}$ in a scheduling slot is
given by
\begin{eqnarray}
Y_{i,k}=\sqrt{P_{i,k}l_k g_k}H_{i,k}X_{i,k} +Z_{i,k},
\end{eqnarray}
where $X_{i,k}$, $P_{i,k}$, and ${H}_{i,k}$  are the transmitted
symbol, the transmitted power, and the  multipath fading coefficient  between
the transmitter and user $k$ on subcarrier $i$, respectively. We assume that the transmitted symbol is zero mean with variance
${\cal E}\{\abs{X_{i,k}}^2\}=1,\,\forall i,k,$ where ${\cal E}\{\cdot\}$ denotes statistical expectation. $l_k$ and $g_k$ represent the
path loss and shadowing between the transmitter and user $k$, respectively.
 $Z_{i,k}$ is the  additive white Gaussian noises (AWGN)  on subcarrier $i$ at user $k$  with zero mean and variance $\sigma_{z}^2$.

\subsection{ Information Decoding and Energy Harvesting Receiver}
\label{sect:receiver}
In this paper, we assume that each user has the ability to decode the modulated information and to harvest energy\footnote{The details
of the energy harvesting process are beyond the scope of this paper and interested readers may refer to  \cite{CN:Shannon_meets_tesla} for a detailed description.} from the received radio signal. However,
 the signal used for decoding of the modulated
information cannot be used for harvesting energy \cite{CN:MIMO_WIPT}. As a result,  at any given scheduling slot, a user can either decode information when it is active (being served by the transmitter) or harvest energy when it is idle,
 but not both concurrently, cf. Figure \ref{fig:system_model}.

\section{Resource Allocation }\label{sect:forumlation}
In this section, we define the  system energy efficiency\footnote{We note that, in the paper,
  a normalized  energy unit is adopted for resource allocation algorithm design, i.e., Joule-per-second. Thus,
  the terms ``power" and ``energy" are used interchangeable.}
 and formulate the corresponding resource allocation problem.

\subsection{Instantaneous Channel Capacity}
\label{subsect:Instaneous_Mutual_information}
%%%%%%%%%%%%%%%%%%%%%%%%%%%%%%%%%%%%%%%%%%%%%%%%%%%
In this subsection, we define the adopted system performance
measure. Given perfect CSI at the
user, the channel capacity in a scheduling slot between  the transmitter and  user $k$ on
subcarrier $i$ with channel bandwidth $W$  is given by
\begin{eqnarray}\label{eqn:cap}
C_{i,k}=W\log_2\Big(1+P_{i,k}\Gamma_{i,k}\Big)\,\,\,\,
\mbox{and}\,\,\,\,
\Gamma_{i,k}=\frac{l_k
g_k\abs{H_{i,k}}^2}{\sigma_{z}^2},
\end{eqnarray}
where $ \Gamma_{i,k}$ is the channel-to-noise ratio (CNR) at user $k$ on subcarrier $i$.  On the other hand,  we assume that   in each scheduling slot one user is selected\footnote{ We note that single user transmission
 in each scheduling slot is commonly used in practical multiuser systems such as Wi-Fi systems.  On the other hand,
 the adopted framework can be generalized to the case where the data of different users is multiplexed  on different subcarriers,
at the expense of a more involved notation.   } for information transfer and served by the transmitter. Concurrently,  the remaining $K-1$ idle users harvest energy from the radio signal emitted by the transmitter. Then, the \emph{system capacity}  in a scheduling slot is defined as the total
 number of
 bits successfully delivered to the selected user (bit-per-second) and is given by
\begin{eqnarray}
 \label{eqn:avg-sys-goodput} && \hspace*{-5mm} U({\cal P}, { \cal S})=\sum_{i=1}^{n_F}\sum_{k=1}^K w_k s_k C_{i,k},
\end{eqnarray}
where ${\cal P}=\{ P_{i,k} \ge 0,\, \forall i,k\}$ and ${\cal S}=\{ s_{k}=\{0,1\}, \forall k\}$ are the power allocation and user selection policies, respectively. $w_k\ge0,\,\forall k,$ is a positive constant which allows the transmitter to give different priorities
to different mobile users and to enforce certain notions of fairness. On the other hand, for designing an energy efficient resource allocation algorithm, we incorporate the total power
dissipation of the system in the optimization objective
function.  To this end, we model the power dissipation (Joule-per-second)  in the system as:
\begin{eqnarray}
 \label{eqn:power_consumption}
\hspace*{-2mm}U_{TP}({\cal P},{\cal S})\hspace*{-2mm}&=&\hspace*{-2mm}P_C+
 \sum_{i=1}^{n_F}\sum_{k=1}^{K}\varepsilon  P_{i,k}s_k- P_H \\
 \hspace*{-4mm}\mbox{where}\label{eqn:Power_harvested_d}\,\,P_H \hspace*{-2mm}&=& \hspace*{-0mm}\underbrace{\sum_{i=1}^{n_F}\sum_{k=1}^{K} P_{i,k}s_k \Big(\sum_{j\neq k}\eta_j\abs{H_{i,j}}^2 l_j g_j\Big) }_{\mbox{Power harvested by idle users}}
\end{eqnarray}
and $j\in\{1,\,\ldots,\,K\}$. $P_C>0$ is the \emph{static circuit power dissipation} of device electronics of the active transceiver  such as mixers, filters, digital-to-analog converters, and is independent of
the actual transmitted power. The total power consumption of the $K-1$  idle users is omitted in (\ref{eqn:power_consumption}) since it is relatively small compared to the power dissipation of the active transceiver pair. The middle term on
 the right hand side of (\ref{eqn:power_consumption}) is the power consumption
 in the power amplifier and $\varepsilon\ge 1$ is a constant which accounts for the power inefficiency
of the power amplifier. For example, if $\varepsilon=5$, then   $5$ Watts are consumed in the power amplifier for every 1 Watt of power radiated in the RF which results in a  power
efficiency of $\frac{1}{\varepsilon}=\frac{1}{5}=20\%$; power that is not converted to a useful signal is dissipated as heat in the power amplifier. On the other hand, the minus sign in front of $P_H$ in (\ref{eqn:power_consumption}) indicates that a
part of the power radiated  by the transmitter can be possibly harvested by the idle users.  Here, we assume that the ability to harvest energy is heterogeneous. Specifically, we define the constant parameter $0\le\eta_j\le1$  used in (\ref{eqn:Power_harvested_d}) as user $j$'s efficiency in harvesting  energy from the received radio signal.
The term $\eta_j\abs{H_{i,j}}^2 l_j g_j$ in (\ref{eqn:Power_harvested_d}) can be interpreted as \emph{frequency selective power transfer efficiency} for transferring power from the transmitter to idle user $j$ on subcarrier $i$.  We note that although $U_{TP}({\cal P},{\cal S})$ can be a negative value mathematically since $\abs{H_{i,j}}^2\in[0,\infty)$, $U_{TP}({\cal P},{\cal S})>0$  always holds in practical communication systems. In particular, since the transmitter is the only   RF  energy source of the idle users, we have
$\sum_{i=1}^{n_F}\sum_{k=1}^{K}\varepsilon  P_{i,k}s_k \ge \sum_{i=1}^{n_F}\sum_{k=1}^{K}  P_{i,k}s_k \ge \sum_{i=1}^{n_F}\sum_{k=1}^{K} P_{i,k}s_k \Big(\sum_{j\neq k}\eta_j\abs{H_{i,j}}^2 l_j g_j\Big)$, where the second inequality is due to the second law of thermodynamics in energy flow from physic.

The \emph{energy efficiency} of the considered  system in a scheduling slot  is defined as the
total number of bits successfully delivered to the selected users per unit energy consumption  (bit-per-Joule)   which is given by
\begin{eqnarray}
 \label{eqn:avg-sys-eff} \hspace*{-8mm}U_{eff}({\cal P},{\cal S})&=&\frac{U_{}({\cal P},{\cal S})}{U_{TP}({\cal P},{\cal S})} .
\end{eqnarray}
%%%%%%%%%%%%%%%%%%%%%%%%%%%%%%%%%%%%%%%%%%%%%%%%%%%%%%%
\subsection{Optimization Problem Formulation}
\label{sect:cross-Layer_formulation}
%%%%%%%%%%%%%%%%%%%%%%%%%%%%%%%%%%%%%%%%%%%%%%%%%%%%%%%

The optimal power allocation policy, ${\cal P}^*$,
 and user selection
policy, ${\cal S}^*$, can be obtained by solving
\label{prob1}
\begin{eqnarray}
\label{eqn:cross-layer}&&\hspace*{15mm} \max_{{\cal P}, {\cal S}
}\,\, U_{eff}({\cal P},{\cal S}) \\
\notag \mbox{s.t.} \hspace*{-5mm}&&\mbox{C1:}\,\, \sum_{i=1}^{n_F} \sum_{j=1}^K P_{i,j}s_j \Big(\eta_k\abs{H_{i,k}}^2l_k g_k\Big)\ge (1-s_k)P_{\min_k}^{req},\, \forall k,\notag\\
&&\mbox{C2:}\,\,\notag\sum_{i=1}^{n_F}\sum_{k=1}^KP_{i,k}s_k\le P_{\max}, \\
&&\notag \mbox{C3:}\,\, \sum_{i=1}^{n_F}\sum_{k=1}^K \varepsilon P_{i,k}s_k + P_C \le P_{PG},\\
&&\notag \mbox{C4:}\,\, \sum_{i=1}^{n_F}\sum_{k=1}^K s_k C_{i,k}\ge R_{\min}\hspace*{8mm}\mbox{C5:}\,\, \sum_{k=1}^K s_k \le 1, \\
&&\notag \mbox{C6:}\,\,s_k\in\{0,1\},\,\forall k,\hspace*{17.6mm} \mbox{C7:}\,\, P_{i,k} \ge 0, \forall i,k.
\end{eqnarray}
Here, $P_{\min_k}^{req}$ in C1 denotes the minimum required
 power transfer for  user $k$ if it is idle.  $P_{\max}$ in C2 is the maximum transmit power allowance to control the amount of out-of-cell
interference.  C3 constrains the total power consumption
of the system to not exceed the maximum power supply from the
power grid, $P_{PG}$.  C4 ensures a minimum required system data rate $R_{\min}$. C6 is a combinatorial constraint on the user selection variables. C5 and C6 are imposed to guarantee that in each scheduling slot at most one user is served by the transmitter.
C7 is the non-negative constraint on the power allocation variables.
%%%%%%%%%%%%%%%%%%%%%%%%%%%%%%%%%%%%%%%%%%%%%%%%%%%%%%% %%%%%%%%%%%%%%%
\section{Solution of the Optimization Problem} \label{sect:solution}

%%%%%%%%%%%%%%%%%%%%%%%%%%%%%%%%%%%%%%%%%%%%%%%%%%%%%%%%%%%%%%%%%%%%%%
The optimization problem in (\ref{eqn:cross-layer}) is a mixed non-convex and combinatorial optimization problem. The non-convexity comes from the objective function which is the ratio of
two functions and  the combinatorial nature comes from the integer constraint for user selection. The first step in solving the considered problem is to simplify the objective function using techniques from nonlinear fractional programming.
%%%%%%%%%%%%%%%%%%%%%%%%%%%%%%%%%%%%%%%%%%%%%%%%%%%%%%%%%%%%%%%%%%%%%%%%%%%%%%%
\subsection{Transformation of the Objective Function} \label{sect:solution_dual_decomposition}

%%%%%%%%%%%%%%%%%%%%%%%%%%%%%%%%%%%%%%%%%%%%%%%%%%%%%%%%%%%%%%%%%%%%%%%%%%%%%%%
For the sake of presentation simplicity, we  define $\mathcal{F}$ as the set of
feasible solutions of the optimization problem in
(\ref{eqn:cross-layer}) and $\{{\cal P},{\cal  S}\}\in\mathcal{F}$.   As the power allocation variables are constrained by C2, C3, and C7, $\mathcal{F}$ is a compact set.  Without loss of generality, we define $q^*$ as the
maximum energy efficiency of the considered system which is given by
\begin{eqnarray}
q^*=\frac{U({\cal P^*},{\cal  S^*})}{U_{TP}({\cal
P^*},{\cal  S^*})}=\max_{{\cal P}, {\cal  S}}\,\frac{U({\cal P},{\cal  S})}{U_{TP}({\cal P},{\cal  S})}.
\end{eqnarray}
We are now ready to introduce the following Theorem which is borrowed from nonlinear fractional programming  \cite{JR:fractional}.
\begin{Thm}\label{Thm:1}
The maximum energy efficiency $q^*$  is achieved if and only if
\begin{eqnarray}\notag \label{eqn:penalty_q}
\max_{{\cal P}, {\cal  S}}&& \hspace*{-2mm}\,U({\cal
P},{\cal  S})-q^*U_{TP}({\cal P},{\cal  S})\\
 =&& \hspace*{-2mm}U({\cal
P^*},{\cal  S^*})-q^*U_{TP}({\cal P^*}, {\cal
 \cal S^*})=0,
\end{eqnarray}
\end{Thm}
for $U({\cal P},{\cal  S})\ge0$ and $U_{TP}({\cal P},{\cal  S})>0$.

 \emph{\,Proof:} Please refer to \cite[Appendix A]{JR:TWC_large_antennas} for a proof of Theorem 1.

Theorem \ref{Thm:1} provides a necessary and sufficient condition for the optimal resource allocation policy.
Specifically,  for an optimization problem
with an objective function in fractional form, there exists an
equivalent optimization problem with an
objective function in subtractive form, e.g. $U({\cal P},{\cal  S})-q^*U_{TP}({\cal P}, {\cal  S})$ in the considered
case,  such  that both problem
formulations lead to the same optimal resource allocation policy.  As a result, we can focus on the equivalent objective function
in the rest of the paper.

\begin{table}[t]\caption{Iterative Resource Allocation Algorithm.}\label{table:algorithm}
\vspace*{-5mm}\small
\begin{algorithm} [H]                    % enter the algorithm environment
\caption{Iterative Resource Allocation Algorithm }          % give the algorithm a caption
\label{alg1}                           % and a label for \ref{} commands later in the document
\begin{algorithmic} [1]
\normalsize           % enter the algorithmic environment
\STATE Initialize the maximum number of iterations $L_{max}$ and the
maximum tolerance $\epsilon$
 \STATE Set maximum energy
efficiency $q=0$ and iteration index $n=0$

\REPEAT [Main Loop]
%%%%%%%%%%%%%%%%%%%%
\STATE Solve the inner loop problem in ($\ref{eqn:inner_loop}$) for
a  given $q$ and obtain resource allocation policy $\{{\cal P'}, {\cal  S'}\}$
%%%%%%%%%%%%%%%%%%%%
\IF {$U({\cal P'}, {\cal  S'})-q U_{TP}({\cal P'},{\cal  S'})<\epsilon$} \STATE  $\mbox{Convergence}=\,$\TRUE \RETURN
$\{{\cal P^*,\cal S^*}\}=\{{\cal P',\cal S'}\}$ and $q^*=\frac{U({\cal
P'},{\cal  S'})}{ U_{TP}({\cal P'}, {\cal S'})}$
 \ELSE \STATE
Set $q=\frac{U({\cal P'}, {\cal  S'})}{ U_{TP}({\cal
P'},{\cal  S'})}$ and $n=n+1$ \STATE  Convergence $=$ \FALSE
 \ENDIF
 \UNTIL{Convergence $=$ \TRUE $\,$or $n=L_{max}$}

\end{algorithmic}
\end{algorithm}\vspace*{-10mm}
\end{table}

\subsection{Iterative Algorithm for Energy Efficiency Maximization}
In this section, an iterative algorithm (known as the
Dinkelbach method \cite{JR:fractional}) is proposed for solving
(\ref{eqn:cross-layer}) with an equivalent objective function such that the obtained resource allocation policy satisfies the conditions stated in Theorem 1. The proposed algorithm is summarized in Table \ref{table:algorithm} and
the convergence to the maximum energy efficiency is guaranteed if the inner problem (\ref{eqn:inner_loop}) in each iteration can be solved.

\emph{Proof: }Please refer to \cite[Appendix B]{JR:TWC_large_antennas}  for a proof of convergence.

As shown in Table \ref{table:algorithm}, in each iteration in the
main loop, i.e., in lines 3--12,  we solve the following optimization problem for a given
parameter $q$:
\begin{eqnarray}\label{eqn:inner_loop}
&&\hspace*{-10mm}\max_{{\cal P},  {\cal  S}}
\,\,\,{U}({\cal P},{\cal  S})-q{U}_{TP}({\cal P},{\cal  S}
)\nonumber\\
&&\hspace*{-15mm}\mbox{s.t.} \,\,\mbox{C1, C2, C3, C4, C5, C6, C7}.
\end{eqnarray}

We note that ${U}({\cal P},{\cal  S})-q{U}_{TP}({\cal P},{\cal  S}
)\ge 0$ holds  for any value of $q$ generated by Algorithm I. Please refer  to \cite[Proposition 3]{JR:TWC_large_antennas} for a proof. On the other hand, it can be observed that
the problem formulation in (\ref{eqn:inner_loop}) for energy efficiency maximization
is a generalized problem formulation for
aggregate weighted system capacity maximization. Indeed, if we set
$q=0$, then the objective function in (\ref{eqn:inner_loop}) will become the aggregate weighted system
capacity.

%%%%%%%%%%%%%%%%%%%%%%%%%%%%%%%%%%%%%%%%%%%%%%%%%%%%%%%
\subsubsection*{Solution of the Main Loop Problem}
The transformed problem  is now a mixed convex  and combinatorial optimization problem. The integer constraint for user selection in C6  is still an obstacle in tackling the problem.  Indeed, the traditional brute force approach or a branch-and-bound method can be used to obtain a global
optimal solution but result in a prohibitively high complexity with respect to (w.r.t.) the numbers of users and subcarriers. In order to strike a balance
between computational complexity and optimality, we follow the approach in
\cite{JR:Roger_OFDMA} and relax $s_k$ in constraint C6
to be a real value between zero and one instead of a Boolean, i.e., $\mbox{C6: } 0\le s_k\le 1,\,\forall k$. Then,
$s_k$ can be interpreted as a time-sharing factor for
the $K$ users to utilize the subcarriers.   For facilitating the time-sharing, we introduce a new variable and define it as
$\tilde{P}_{i,k}={P}_{i,k}s_{k},\, \forall i,k$. The variable
represents the actual transmitted power in the RF of the transmitter on subcarrier
$i$ for user $k$ under the time-sharing assumption. Although the relaxation of the user selection constraint
will generally lead to a suboptimal solution, the authors in
\cite{JR:limited_backhaul}  show that the duality gap (suboptimality)
due to the constraint relaxation becomes zero when the number of subcarriers
is sufficiently large, e.g. $n_F=128$.

With this relaxation, it can be shown that the problem is now jointly concave w.r.t. the power allocation and user selection variables under the time-sharing assumption.  As a result, under some mild conditions, solving the dual
problem is equivalent to solving the primal problem \cite{book:convex}.

 \subsection{Dual Problem Formulation}
In this subsection, we solve the resource allocation
optimization problem by solving its dual for a given value of $q$.
For this purpose, we  need the Lagrangian function of the primal problem in (\ref{eqn:inner_loop}) which is given by
\begin{eqnarray}\hspace*{-2mm}&&{\cal
L}( \boldmath\textbf{$\alpha$}, \beta,\gamma,\lambda,\delta,{\cal P},{\cal  S})\\
\notag\hspace*{-7mm}&=&\hspace*{-2mm}\sum_{k=1}^K\sum_{i=1}^{n_F}
(w_k+\gamma)s_k C_{i,k}\hspace*{-0.5mm}-\hspace*{-0.5mm}q\Big(U_{TP}({\cal P},{\cal S})\Big)\\
\hspace*{-6mm}&-&\hspace*{-2mm}\lambda\Big(\sum_{i=1}^{n_F}\sum_{k=1}^K \varepsilon \tilde{P}_{i,k}+ P_C - P_{PG}\Big)-\delta\Big(\sum_{k=1}^K s_k - 1 \Big)\notag\\
\hspace*{-6mm}&-&\hspace*{-2mm}\beta\Big(\sum_{i=1}^{n_F}\sum_{k=1}^K \tilde{P}_{i,k}s_k- P_{\max}\Big)-\gamma R_{\min}\notag\\
\hspace*{-6mm}&-&\hspace*{-2mm}\sum_{k=1}^K\alpha_k\Big((1-s_k)P_{\min_k}^{req}-\sum_{i=1}^{n_F} \sum_{j=1}^K \tilde{P}_{i,j} \Big(\eta_k\abs{H_{i,k}}^2l_k g_k\Big) \Big).\notag
\label{eqn:Lagrangian}
\end{eqnarray}
Here, $\boldmath\textbf{$\alpha$}$ has elements $\alpha_k$, $k\in\{1,\ldots,K\}$,
 and is the Lagrange multiplier vector accounting for the minimum required power transfer for idle users in C1. $\beta\ge0$ is the Lagrange multiplier
corresponding to the maximum transmit power limit in C2.  $\lambda\ge0$ is the Lagrange multiplier for C3 accounting for the maximum  power dissipation in the transmitter due to the limited power supply from the power grid.
$\gamma\ge0$ and $\delta\ge 0$ are the Lagrange multipliers associated with the minimum data rate requirement and the user selection constraint in C4 and C5, respectively.  On the other hand, the boundary constraints C6
and C7 on the user selection  and  power variables will be absorbed into the Karush-Kuhn-Tucker (KKT) conditions
when deriving the optimal resource allocation policy in the
following.

Thus, the dual problem
 is given by
\begin{eqnarray}
\underset{ \boldmath\textbf{$\alpha$}, \beta,\gamma,\lambda,\delta \ge 0}{\min}\ \underset{{\cal
P,\cal S}}{\max}\quad{\cal
L}( \boldmath\textbf{$\alpha$}, \beta,\gamma,\lambda,\delta,{\cal P},{\cal  S}).\label{eqn:master_problem}
\end{eqnarray}

\subsection{Lagrange Dual Decomposition }
\label{sect:sub_problem_solution} By Lagrange dual decomposition, the dual
problem can be decomposed into two layers: Layer 1 (inner maximization in (\ref{eqn:master_problem})) consists of $n_F+1$ subproblems where $n_F$ of them have identical structure and can be solved in parallel; Layer 2 (outer minimization in (\ref{eqn:master_problem})) is the master problem. The dual problem can be
solved by solving the problems in Layer 1 and Layer 2 iteratively, where in each iteration, the transmitter solves the
 subproblems  by using the KKT conditions for a fixed set of Lagrange multipliers, and the master problem  is solved using the gradient method.

\subsubsection*{Layer 1 (Subproblem Solution)}
 Using standard
optimization techniques and the KKT conditions, the closed-form optimal power allocation
on subcarrier $i$ for user $k$ for a given $q$  is obtained as
 \begin{eqnarray}\label{eqn:power1}
\tilde{P}_{i,k}^*\hspace*{-2.5mm}&=&\hspace*{-2.5mm}s_k{P}_{i,k}^*=s_k\Bigg[\frac{W(w_k+\gamma)}{\ln(2)\Theta_{i,k}}-\frac{1}{\Gamma_{i,k}}\Bigg]^+\hspace*{-1.5mm}, \,\forall i,k,\,\\
\hspace*{-3mm}\mbox{where  }\Theta_{i,k}\hspace*{-2.5mm}&=&\hspace*{-2.5mm}q\Big(\varepsilon-\sum_{j\ne k}\eta_j g_j l_j\abs{H_{i,j}}^2 \Big)+\lambda\varepsilon+\beta \notag\\
&&\hspace*{-2.5mm}-\sum_{j\ne k}\alpha_j\eta_j g_j l_j\abs{H_{i,j}}^2
\end{eqnarray}
and $\big[x\big]^+=\max\{0,x\}$.  The optimal power allocation solution in (\ref{eqn:power1}) has the form of multilevel water-filling. In particular, the water-level,  i.e., $\frac{W(w_k+\gamma)}{\ln(2)\Theta_{i,k}}$, is different across different subcarriers and different users. In fact, the water-level on subcarrier $i$ for user $k$ depends not only on the priority of user $k$ via $w_k$, but also on its influence on the other $K-1$ users via $\sum_{j\ne k}\eta_j g_j l_j\abs{H_{i,j}}^2$.
Besides, Lagrange multipliers $\gamma$ and $\alpha_j$
force the transmitter to transmit with a sufficient amount of power to fulfill the system data rate requirement
 $R_{\min}$ and the minimum power transfer requirement $P_{\min_j}^{req}$ for idle user $j$, respectively.

On the other hand, in order to obtain the optimal user selection, we
take the derivative of the subproblem w.r.t. $s_{k}$, which yields $\frac{\partial {\cal
L}( \boldmath\textbf{$\alpha$}, \beta,\gamma,\lambda,\delta,{\cal P},{\cal  S})}{\partial
s_{k}^*}=Q_{k}-\delta+\alpha_k P_{\min_k}^{req}$, where $Q_{k}$ is the marginal benefit achieved by the system
 by selecting user $k$. From (\ref{eqn:Lagrangian}) we obtain $Q_{k}=$
\begin{eqnarray} \label{eqn:subcarrier_allocation}
W(w_k+\gamma)\sum_{i=1}^{n_F}\Bigg(\log_2\Big(\hspace*{-0.5mm}1+P^*_{i,k}\Gamma_{i,k}\Big)-\frac{\Gamma_{i,k}P^*_{i,k}/\ln(2)}{1+{P}^*_{i,k}\Gamma_{i,k}}\hspace*{-0.5mm}\Bigg).
\end{eqnarray}  Thus, the optimal user selection
 is given by
\begin{eqnarray}\hspace*{-1mm}
\label{eqn:sub_selection}s_{k}^*=
 \left\{ \begin{array}{rl}
 1 &\mbox{if $Q_{k}[j]\ge 0$}  \\
 0 &\mbox{ otherwise}
       \end{array} \right. .
\end{eqnarray}
We note that
although $P_C$  does not appear in (\ref{eqn:power1})-(\ref{eqn:sub_selection}), it has an influence
on the solution of the dual problem via the updating process of $q$, cf. Table I.

%
% \begin{figure}[t]
%\centering
%\includegraphics[width=3.5 in]{layer.eps}
% \caption{ An illustration of dual decomposition of a
%large-scale problem into a two-layer problem. } \label{fig:layer}
%\end{figure}

\subsubsection*{Layer 2 (Master Problem Solution)}
For solving the Layer 2 master minimization problem in (\ref{eqn:master_problem}), i.e, to find $\textbf{\boldmath$\alpha$}, \beta,\gamma,\lambda$, and $\delta$ for given $\cal P$
and $\cal S$, the gradient method can be used since the dual function is differentiable. The gradient update
equations are given by:
\begin{eqnarray}\label{eqn:multipler1}\notag
\hspace*{-1.0mm}\alpha_k(m+1)\hspace*{-3.25mm}&=&\hspace*{-3.25mm}\Big[\alpha_k(m)-\xi_1(m)\hspace*{-0.5mm}\times\hspace*{-0.5mm} \Big(\hspace*{-0.5mm}\sum_{i=1}^{n_F}\hspace*{-0.5mm} \sum_{j=1}^K P^*_{i,j}s^*_j \Big(\hspace*{-0.5mm}\eta_k \abs{H_{i,k}}^2l_k g_k\hspace*{-0.5mm}\Big) \Big.\\
\hspace*{-5.0mm}&& \hspace*{-5.5mm} -(1-s^*_k)P_{\min_k}^{req}\Big)\Big]^+\hspace*{-1.5mm},\forall k,\\
\hspace*{-5.0mm}\beta(m+1)\hspace*{-3.25mm}&=&\hspace*{-3.25mm}\Big[\beta(m)\hspace*{-0.5mm}-\hspace*{-0.5mm}\xi_2(m)\hspace*{-0.5mm}\times\hspace*{-0.5mm}
\Big( P_{\max}\hspace*{-0.5mm}-\hspace*{-0.5mm}\sum_{i=1}^{n_F}\sum_{k=1}^KP^*_{i,k}s^*_k\Big)\Big]^+\hspace*{-1.5mm}, \label{eqn:multipler2}\\
\hspace*{-5.0mm}\gamma(m+1)\hspace*{-3.25mm}&=&\hspace*{-3.25mm}\Big[\gamma(m)-\xi_3(m)\hspace*{-0.5mm}\times\hspace*{-0.5mm}
\Big(\sum_{i=1}^{n_F}\sum_{k=1}^K s^*_k C_{i,k}\hspace*{-0.5mm}-\hspace*{-0.5mm} R_{\min}\Big )\Big]^+\hspace*{-1.5mm},\label{eqn:multipler3}\\
\hspace*{-5.0mm}\notag\lambda(m+1)\hspace*{-3.25mm}&=&\hspace*{-3.25mm}\Big[\lambda(m)-\xi_4(m)\hspace*{-0.5mm}\times\hspace*{-0.5mm}
\Big(P_{PG}\Big.\\
 \Big.\hspace*{-5.0mm}&& \hspace*{-5.5mm} -P_C-\sum_{i=1}^{n_F}\sum_{k=1}^K\varepsilon s^*_{k}P^*_{i,k} \Big)\Big]^+\hspace*{-2.2mm}, \label{eqn:multipler4}\\
\hspace*{-5.0mm}\delta(m+1)\hspace*{-3.25mm}&=&\hspace*{-3.25mm}\Big[\delta(m)-\xi_5(m)\hspace*{-0.5mm}\times\hspace*{-0.5mm}
\Big(1- \sum_{k=1}^K s^*_k\Big)\Big]^+\hspace*{-1.5mm}, \label{eqn:multipler5}
\end{eqnarray}
where index $m\ge 0$ is the iteration index  and $\xi_u(m)$,
$u\in\{1,\ldots,5\}$, are positive step sizes.  The updated Lagrange multipliers in
(\ref{eqn:multipler1})--(\ref{eqn:multipler5}) are used for solving
the subproblems in (\ref{eqn:master_problem}) via updating the resource allocation policy in (\ref{eqn:power1})--(\ref{eqn:sub_selection}). Since the transformed problem is concave
for a given parameter $q$, it is
guaranteed that the iteration between Layer 2 (master problem) and Layer 1 (subproblems) converges
to the primal optimum of (\ref{eqn:inner_loop}) in the
main loop, if the chosen step
 sizes satisfy the infinite travel condition
 \cite{book:convex}.

To summarize the iterative algorithm between Layer 1 and Layer 2, the gradient update in (\ref{eqn:multipler1})--(\ref{eqn:multipler5}) can be interpreted as the
pricing adjustment rule of the demand and supply model \cite{book:convex}. Specifically, the Lagrange multipliers can be interpreted as a set of shadow prices for utilizing the resources. If the demand of the resource exceeds
the supply, then the gradient method will raise
the shadow prices via adjusting  the Lagrange  multipliers  in the next iteration; otherwise, it will reduce the
shadow prices until some users can afford them. By combining
the gradient update equations and the user selection criterion in (\ref{eqn:sub_selection}), only one user is   selected eventually even though  time-sharing is introduced for solving the transformed problem in (\ref{eqn:Lagrangian}).

\begin{figure}[t]\vspace*{-8mm}
 \centering
\includegraphics[width=3.5 in]{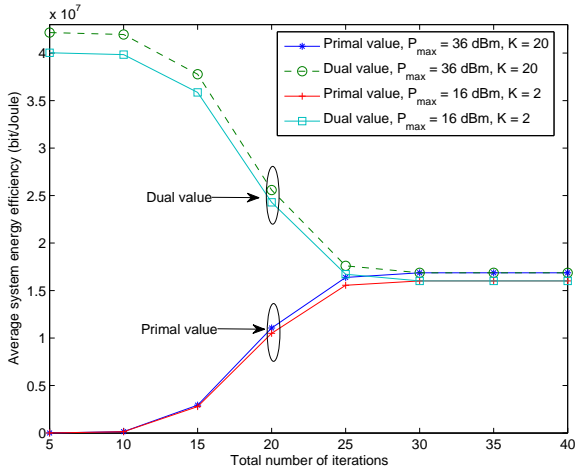}\vspace*{-4mm}
\caption{Average system energy efficiency (bit-per-Joule) versus number of
iterations with different
 numbers of users, $K$, and different
values of maximum transmit power allowance, $P_{\max}$.  } \label{fig:convergence}\vspace*{-6mm}
\end{figure}
\vspace*{-1mm}
\section{Simulation Results}\vspace*{-0.5mm}
\label{sect:result-discussion} In this section, we evaluate the
 performance of the proposed resource allocation algorithm using simulations. An indoor communication system  with a maximum service distance
  of 10 meters is considered.  The TGn path loss model \cite{report:tgn} is adopted with a reference distance of $d_0=2$ meters. The $K$ users are uniformly distributed between the reference distance and the maximum service distance. The effective antenna gain for each
   transceiver is assumed to be $14$ dB.   The system bandwidth is ${\cal B}=5$ MHz, the number of subcarriers is $n_F=128$, and $w_k=1,\,\forall k$.  Note
that by setting $w_k=1,\,\forall k$, we obtain the  maximum achievable
system energy efficiency. We assume  a carrier
center frequency of $470$ MHz which will be used by IEEE 802.11 for the next generation of Wi-Fi systems  \cite{report:80211af}.
Each subcarrier for RF transmission has a bandwidth of $W=39$ kHz and the  noise
variance is $\sigma_{z}^2=-118$ dBm.  The multipath fading coefficients of
the transmitter--user links are generated as independent and identically
distributed (i.i.d.)  Rayleigh  random variables  with unit variance.  The shadowing of all communication links is set to $0$ dB, i.e., $g_k=1,\,\forall k$.
We assume a static circuit power
consumption of $P_C=$ 40 dBm, a maximum power grid supply of $P_{PG}=50$ dBm,  and a minimum data rate
requirement of $R_{\min}=10$ Mbits/s. The  minimum required power transfer and the energy harvesting efficiency  are set to $P_{\min_k}^{req}=-10$ dBm,\, $\forall k$, and  $\eta_k=0.8,\, \forall k,$ respectively.
Besides, we assume a power efficiency of
$40\%$ for the power amplifier used at the transmitter,
i.e., $\varepsilon=\frac{1}{0.4}=2.5$. The average  energy efficiency is computed according to (\ref{eqn:avg-sys-eff}) and averaged over multipath  fading and path loss.  In the  sequel, the total number of iterations is defined as the number of main loops  in Algorithm 1 multiplied with the number
of iterations in solving the Layer 1 and Layer 2 problems. Moreover, the step sizes adopted in (\ref{eqn:multipler1})--(\ref{eqn:multipler5}) are optimized for obtaining a fast convergence. Note that if the transmitter is
unable to fulfill  the minimum required system data rate $R_{\min}$ or
the minimum required power transfer $P^{req}_{\min_k}$, we set the energy
efficiency and the system capacity for that channel realization
to zero to account for the corresponding failure.

\subsection{Convergence of Iterative Algorithm 1 }
%%%%%%%%%%%%%%%%%%%%%%%%%%%%%%%%%%%%%%%%%%%%%%%%%%%%%
Figure \ref{fig:convergence} illustrates the convergence behavior and the duality gap of  the
proposed iterative algorithm for maximizing the system energy efficiency.  The duality gap is defined as the difference between primal optimum and dual optimum (achieved by the proposed algorithm) and indicates the suboptimality due to the constraint relaxation in C6 and insufficient numbers of iterations. We investigate the system performance for different numbers of users, $K$,  and different values for the maximum transmit power allowance, $P_{\max}$.
  The results in Figure
\ref{fig:convergence} were averaged over $10^5$ independent
 channel realizations for both path loss and multipath fading. It can be observed that the energy efficiency of the proposed algorithm converges to the
optimum value within 30 iterations for all considered scenarios. The fast convergence and the zero duality gap  confirm the practicality of  the proposed algorithm.

In the following results, we set the total number of iterations to 30 to illustrate the performance of the proposed algorithm.

\begin{figure}[t]\vspace*{-8mm}
 \centering
\includegraphics[width=3.5 in]{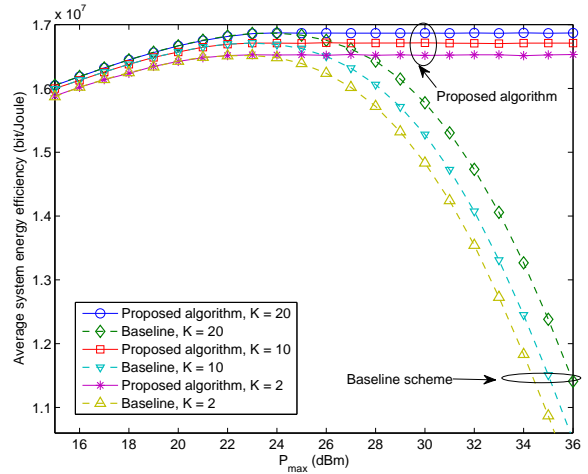}\vspace*{-4mm}
\caption{Average system energy efficiency (bit-per-Joule) versus
maximum transmit power allowance, $P_{\max}$, for different numbers of users, $K$. } \label{fig:EE_PT}\vspace*{-6mm}
\end{figure}

\vspace*{-1mm}
\subsection{Average Energy Efficiency and Average System Capacity }\vspace*{-0.5mm}
Figure \ref{fig:EE_PT} depicts the  average system energy efficiency versus the
maximum transmit power allowance, $P_{\max}$, for different numbers of users, $K$.
 It can be
observed that   the energy efficiency of the proposed algorithm increases w.r.t. $P_{\max}$ monotonically  and  reaches an upper limit where the energy efficiency gain due to a higher value of $P_{\max}$ vanishes. This result indicates that once the maximum energy efficiency is achieved by transmitting a sufficiently large power in the RF, any additional increase in the transmitted power will incur  a loss in energy efficiency. On the other hand, the energy efficiency of the system increases with the number of users and there are two
 reasons for this behaviour. First, as the number of users in the system increases, the transmitter has a higher chance
to select a user who has a strong channel due to multiuser diversity (MUD). Indeed, the MUD introduces an extra power gain to the system which helps save
 energy. Second, more idle users harvest the  power radiated by the transmitter which reduces the total power consumption of the system, cf. (\ref{eqn:power_consumption}).  For comparison, Figure
\ref{fig:EE_PT} also contains the energy efficiency of a baseline scheme which adopts a
resource allocation algorithm maximizing the system
capacity (bit/s) under constraints C1--C7. It can be seen
that in the low transmit power allowance regime, the proposed algorithm performs virtually the same as the  baseline scheme. Indeed, the small power radiated by the transmitter creates a bottleneck in the system and
the performance of the proposed algorithm is restricted by the limited system resources.  However, in the high transmit power allowance
regime,  the energy efficiency of the
baseline scheme decreases dramatically since an exceedingly large transmit power is used for  capacity maximization.

\begin{figure}[t]\vspace*{-8mm}
 \centering
\includegraphics[width=3.5 in]{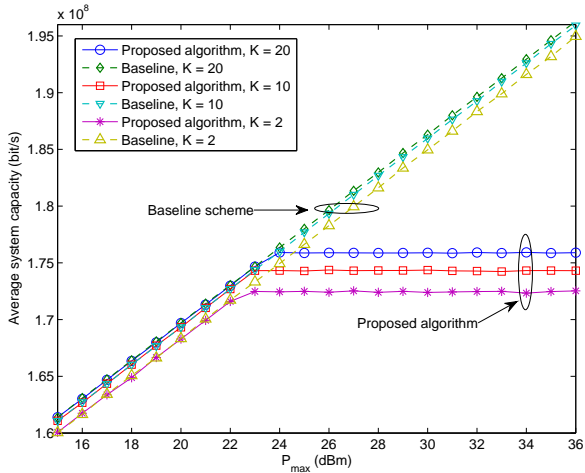}\vspace*{-4mm}
\caption{Average system capacity (bit-per-second) versus
maximum transmit power allowance, $P_{\max}$, for different numbers of users, $K$. } \label{fig:cap_pt}\vspace*{-6mm}\vspace*{-0.5mm}
\end{figure}

Figure \ref{fig:cap_pt} shows the average system capacity versus
maximum transmit power allowance, $P_{\max}$, for different numbers of users, $K$.  The proposed algorithm is compared with the baseline scheme described in the last section.
It can be observed that both schemes benefit from an increasing number of users due to MUD. On the other hand, the proposed
algorithm achieves virtually the same system capacity as the baseline scheme in the low transmit power regime. This suggests that the proposed algorithm transmits with full power in the low transmit power allowance regime. However, as the transmit power allowance
 increases, the baseline scheme outperforms the proposed algorithm,
since the former scheme uses all the available power for  capacity maximization which impairs the system energy efficiency.

\vspace*{-2mm}
\subsection{Average Harvested Power}
\begin{figure}[t]\vspace*{-8mm}
\centering
\includegraphics[width=3.5 in]{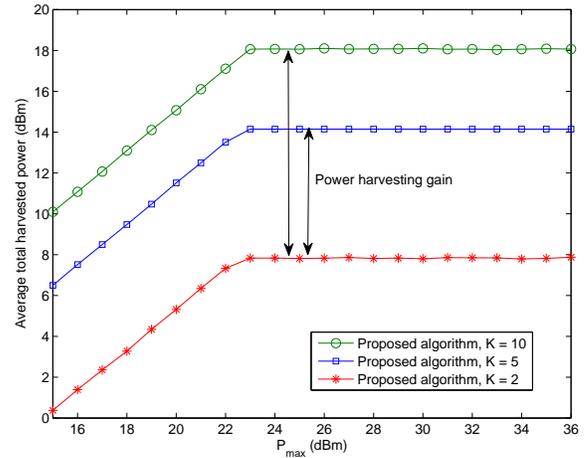}\vspace*{-4mm}
 \caption{Average total harvested power  versus
maximum transmit power allowance, $P_{\max}$, for different numbers of users, $K$. The double-sided arrows indicate the power harvesting gain due to an increasing number of users.} \label{fig:PT_PT}\vspace*{-6mm}\vspace*{-0.5mm}
\end{figure}

%%%%%%%%%%%%%%%%%%%%%%%%%%%%%%%%%%%%%%%%%%%%%%%%%%%%%%%%%%%%%%%%%%%%%%%%%%%%%%%%%%%%%%%%%%
Figure \ref{fig:PT_PT} shows  the  total average  power harvested by the idle users versus the
maximum transmit power allowance, $P_{\max}$, for different numbers of users, $K$.
It can be observed in Figure \ref{fig:PT_PT} that the total average  harvested power increases with $P_{\max}$ since more power  is available in the RF. On the other hand,  a larger portion  of the radiated power  can be harvested  when  there are more users in the system
 since more idle users are participating in the energy harvesting process.

\section{Conclusions}\label{sect:conclusion}\vspace*{-0.5mm}
In this paper, we formulated the resource allocation algorithm design for multiuser OFDM systems as a non-convex and combinatorial optimization problem, in which concurrent wireless information and power transfer were considered. By exploiting nonlinear fractional programming and Lagrange dual decomposition, a novel iterative resource allocation algorithm was proposed
for maximizing the system energy efficiency.  Simulation results showed
that the proposed algorithm converges within a small number
of iterations and unveiled the potential benefits of  having multiple users for energy efficiency, system capacity, and wireless power transfer.

\bibliographystyle{IEEEtran}
\bibliography{OFDMA-AF}

\end{document}